\documentclass[aps,prl,reprint,superscriptaddress,showpacs,amsmath,amssymb]{revtex4-1}

\usepackage{esint}
\usepackage{amsmath,amssymb,amsfonts}
\usepackage{ulem}
\usepackage{graphicx}
\usepackage{epstopdf}
\usepackage{xcolor}
\DeclareGraphicsExtensions{.eps}

\definecolor{darkblue}{rgb}{0, 0, 0.8}
\usepackage[colorlinks=true, breaklinks=true, linkcolor=darkblue, citecolor=darkblue, urlcolor=darkblue]{hyperref}
\newcommand{\doilink}[2]{\href{http://dx.doi.org/#1}{#2}}

\newcommand{\beq}{\begin{equation}}
\newcommand{\eeq}{\end{equation}}

\begin{document}
\title{Near-resonant light scattering by a sub-wavelength ensemble of identical atoms}
\date{\today}
\author{N.J. Schilder}
\affiliation{Laboratoire Charles Fabry, Institut d'Optique Graduate School, CNRS,
Universit\'e Paris-Saclay, F-91127 Palaiseau Cedex, France.}
\author{C. Sauvan}
\affiliation{Laboratoire Charles Fabry, Institut d'Optique Graduate School, CNRS,
Universit\'e Paris-Saclay, F-91127 Palaiseau Cedex, France.}
\author{Y.R.P. Sortais}
\affiliation{Laboratoire Charles Fabry, Institut d'Optique Graduate School, CNRS,
Universit\'e Paris-Saclay, F-91127 Palaiseau Cedex, France.}
\author{A. Browaeys} 
\affiliation{Laboratoire Charles Fabry, Institut d'Optique Graduate School, CNRS,
Universit\'e Paris-Saclay, F-91127 Palaiseau Cedex, France.}
\author{J.-J. Greffet}
\affiliation{Laboratoire Charles Fabry, Institut d'Optique Graduate School, CNRS,
Universit\'e Paris-Saclay, F-91127 Palaiseau Cedex, France.}

\begin{abstract}
We study theoretically the scattering of light by an ensemble of $N$ resonant atoms in a sub-wavelength volume. 
We consider the low intensity regime so that each atom responds linearly to the field. 
While $N$ non-interacting atoms would scatter $N^2$ more than a single atom, 
we find that $N$ interacting atoms scatter less than a single atom near resonance. 
In addition, the scattered power presents strong fluctuations,
either from one realization to another or when varying the excitation frequency. 
We analyze this counter-intuitive behavior  in terms of collective modes resulting from the
light-induced dipole-dipole interactions. We find that for small samples, their properties 
are governed only by their volume when $N\gtrsim 20$. 
\end{abstract}
\maketitle
\vspace{5mm}

When deriving the optical properties of bulk materials, one usually starts from a microscopic
description of the medium as a collection of atoms featuring, in the simplest case, 
one resonance with frequency $\omega_0=2\pi c/\lambda$ and linewidth $\Gamma_0$~\cite{Jackson,BornWolf}.  
When the medium is dilute, such as in a vapor, the field scattered by the 
atomic ensemble is simply the interference of the fields scattered by the individual atoms
considered as {\it independent}, i.e. non-interacting. 
For larger densities, the interactions between the light-induced atomic dipoles play an increasing role
and modify the scattering. 
This raises the question of what happens if near-resonant light is scattered from an atomic sample
with sub-wavelength size. 
By analogy with a resonant dipolar nano-particle~\cite{Novotny2006}, 
one would expect that the scattering cross section $\sigma_{\rm sc}$ is on the order of $\lambda^2$
and independent of the exact number of atoms in the ensemble when it is dense enough.
But how do we reconcile this intuition with the fact that 
for a sub-wavelength-sized sample the fields scattered 
by the $N$ atoms should be in phase and hence 
the cross section should vary like $N^2$? 

The answer to this question relies on a description of the atomic sample in terms of collective modes 
resulting from the interactions between the dipoles induced by the driving light field
\cite{Pierrat2010,Chomaz2012,Bettles2015,Kupriyanov2013,Li2013,Guerin2016,Zhu2016}\cite{footnote_hotvapor}. 
As analyzed by many authors, each of the collective modes has an eigen-frequency and a natural linewidth, 
which depend crucially on the exact spatial arrangement of the atoms~
\cite{Bellando2014,Goetschy2011,Skipetrov2011,Asenjo-Garcia2017}. 
We have shown in a previous work~\cite{Schilder2016} that
for atomic samples with a size $L$ on the order of a few $\lambda$, a few dominant modes
delocalized over the entire system, called
polaritonic modes, play an important role in the scattering of light. 
They correspond to the electromagnetic 
modes found when the ensemble of atoms is described by an effective dielectric constant. 
In a following work investigating the homogenization of these systems~\cite{Schilder2017}, we have also found that
the total power scattered close to resonance saturates when increasing the atom number. 
However, for $L$ around a few $\lambda$, we could not find a simple expression for the saturation value. 

Here we show that for a sub-wavelength volume ($L\ll \lambda$), for which no polaritonic mode exists, 
the strong light-induced interactions between atoms lead to a
scattering cross section averaged over the excitation spectrum that is actually {\it smaller} than 
the one of a single atom, {\it i.e} the sample scatters less than a single atom! 
Furthermore, the cross section presents large fluctuations as a function of the laser frequency. 
These two properties are independent of the atom number for $N\gtrsim 20$. 
This behavior is at odds with the two naive pictures mentioned above: 
a particle with an effective refractive index and $N$ atoms scattering coherently. 
The question of the optical properties of sub-wavelength-sized ensemble
was raised by Dicke~\cite{Dicke1954} for the case where all the atoms are initially excited. 
It was realized later that the interactions suppress the 
resulting super-radiant emission~\cite{Friedberg1972,Haroche1982}. 
Here instead, we concentrate on the scattering in the low light intensity limit where the
atomic dipoles respond linearly to the field. 

\begin{figure}
\begin{centering}
\includegraphics[width=\linewidth]{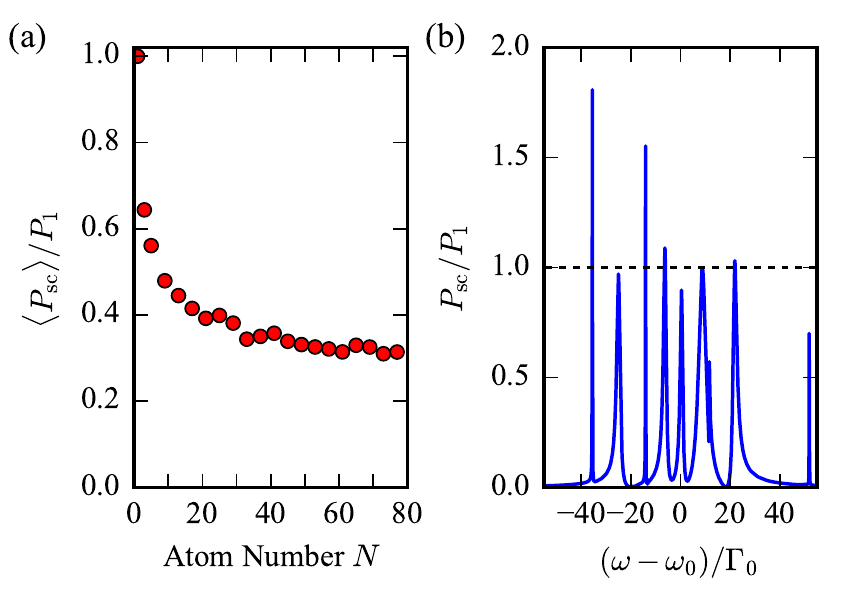}
\caption{(a) Configuration-averaged power scattered by a cubic box 
with sides $\lambda/(2\pi)$ ($Vk^3=1$) containing $N$ atoms illuminated
by a plane wave with frequency $\omega=\omega_0$, 
normalized by the power $P_1$ scattered by a single atom on resonance. 
(b) Normalized scattered power for a single realization of the box with $N=10$
as a function of the light frequency $\omega$. }
\label{Fig1}
\end{centering}
\end{figure}

To investigate the scattering by a sub-wavelength ensemble, 
we apply the model developed in our previous 
works~\cite{Schilder2016,Schilder2017} to the case of a cubic 
box with sides $\sim\lambda/(2\pi)=1/k$ containing $N$ identical atoms at rest.
We consider each atom as a resonant scatterer 
characterized by a classical polarizability $\alpha$. For the sake of simplicity,
we assume that the dipoles can only oscillate linearly along the $z$-axis~\cite{Footnote:2lvl}. 
The wavelength of the atomic transition is $\lambda=2\pi c/\omega_0=780$ nm, 
and the spectral width  $\Gamma_0=2\pi \times 6$ MHz (case of rubidium D2 line relevant 
for experiments~\cite{Bienaime2010,Bender2010,Pellegrino2014,
Jennewein2016,Roof2016,Guerin2016b,Corman2017,Jennewein2018}).
We illuminate the system with a monochromatic 
plane wave propagating along the $x$-axis and linearly polarized along the $z$-axis. 
Each atom is driven by the incident  light field (frequency $\omega$) and the sum of the $z$-components 
of the fields scattered by all other atomic dipoles. 
This leads to a set of coupled dipole equations from which we calculate, in steady
state, the induced dipoles. We finally compute the total scattered field 
$\textbf{E}_{\rm sc}$ and the total scattered power evaluated on a 
spherical surface $\Sigma$ in the far field:
\beq\label{Eq:Pcoh}
P_{\rm sc}=\frac{\varepsilon_0c}{2}\oint_{\Sigma} |\textbf{E}_{\text{sc}}|^2\text{d}S.
\eeq
We generate 1000 realizations of this system with $N$ dipoles  positioned in the volume
according to a uniform random distribution
and calculate the ensemble-averaged total scattered power $\langle P_{\rm sc}\rangle$. 
Figure~\ref{Fig1}(a) shows $\langle P_{\rm sc}\rangle$, normalized by the total power $P_1$
scattered by a single atom on resonance, as a function of the number of particles $N$ 
in a fixed volume $V=k^{-3}$. 
Since the sample is smaller than $\lambda$ in all dimensions, a model based on 
{\it single scattering}, {\it i.e.} neglecting the interactions between dipoles, 
would predict a scattered field $\propto N$, 
leading to a scattered power enhanced 
by a factor $N^2$ as compared to a single atom~\cite{Jackson,Bohren,Berkeley}. 
Surprisingly, we find that $N$ dipoles inside the box 
scatter less than a single dipole. We also find that the scattered 
power is almost independent of the number of dipoles for  $N\gtrsim 20$. 
 
To analyze this behavior, we plot in Fig.~\ref{Fig1}(b) the scattered power as a function 
of the incident frequency $\omega$ for a single realization of the
spatial atomic distribution ($N=10$). 
If we had ignored the interactions between the light-induced dipoles, all modes would be 
degenerate at $\omega_0$ with spectral width $\Gamma_0$. Including the interactions,  
the single atom  Lorentzian spectrum is replaced by a 
scattering spectrum displaying several resonances which are the signature of collective modes. 
This is a consequence of the dipole-dipole interaction, which lifts the degeneracy of the atomic modes 
and generates $N$ spectrally separated collective modes. 
The fact that the peak normalized scattered power $P_{\rm sc}/P_1$ is almost constant and close to 1 
for all modes  indicates that their cross-section is close to 
$3\lambda^2/(2\pi)$, the universal resonant  cross section of a non-lossy dipolar mode. 
This is not surprising as the system size is  smaller than $\lambda$ so that the scattering 
is  dominated by dipolar modes. 
A few modes have a larger scattering cross section and small spectral 
width, which indicates that they are higher order modes resulting from the finite size of the ensemble. 
Indeed, the maximum scattering cross section of a $l$-order mode varies as 
$\sigma_l\sim (2l+1)\lambda^2/(2\pi)$ \cite{Bohren,Miroshnichenko2018}, with 
the dipolar case corresponding to $l=1$, the quadrupolar mode to $l=2$, etc.  
We can now understand why the ensemble-averaged 
scattering cross section of the ensemble is smaller than $3\lambda^2/(2\pi)$: 
due to the spectral separation between the modes, 
a laser line centered on the atomic resonance ($\omega=\omega_0$)
excites mainly one mode with scattering cross section $\sim3\lambda^2/(2\pi)$ 
or misses the modes. As their frequencies vary from one realization 
of the spatial distribution to another,  the ensemble-averaged  
cross-section  is smaller than $3\lambda^2/(2\pi)$.

\begin{figure}
\includegraphics[width=\linewidth]{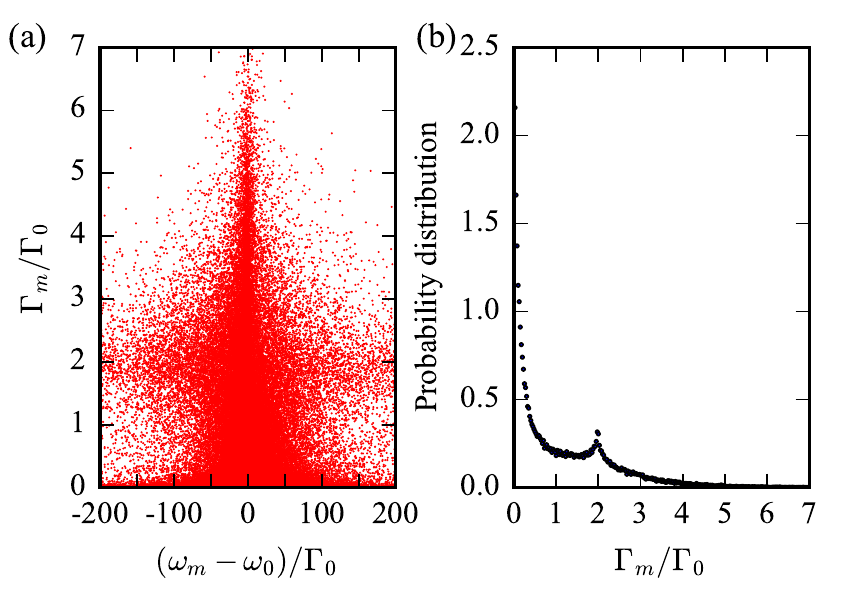}
\caption{Distribution of the complex eigenvalues of the collective modes 
for $N=10$ dipoles placed in a volume of size $Vk^3=1$. 
(a)  Linewidth $\Gamma_m$   versus frequencies $\omega_m$
of the modes calculated for 1000 realizations. 
(b) Normalized probability distribution of the collective linewidths extracted from Fig.~\ref{Fig2}(a), 
corresponding  to 10000 realizations.}
\label{Fig2}
\end{figure}

We now turn to the discussion of the spectral widths $\Gamma_m$ of the modes. 
Figure~\ref{Fig2}(a) shows the distribution of frequencies $\omega_m$ and widths $\Gamma_m$ 
of the collective modes obtained from 1000 realizations of the cloud for $N=10$ dipoles and $Vk^3=1$.
The widths vary significantly, indicating the presence of superradiant (broad) modes 
and subradiant (narrow) modes. However, their spectral widths are linked by a sum rule. 
To see that, we write the set of homogeneous linear equations describing the modes of the system:  
$[{\bf A}]\textbf{P} = (\omega_m -i \frac{\Gamma_m}{2})\textbf{P}$ 
where $\textbf{P}$ is the vector containing the dipole moments of all the atoms and 
$[{\bf A}]$ is the matrix connecting them:
\[ [{\bf A}]=\left( \begin{array}{ccccccc}
&(\omega_0-i\frac{\Gamma_0}{2})&& &
-\frac{3\pi c\Gamma_0}{\omega_0}G_{21}&&\hdots \\
&&&&&&\\
&-\frac{3\pi c\Gamma_0}{\omega_0} G_{12}&&&
(\omega_0-i\frac{\Gamma_0}{2})&&\\
&\vdots&&&&&\ddots\end{array} \right).\] 
Here, $G_{ij}=[{\bf G}]_{xx}(\textbf{r}_j,\textbf{r}_i;\omega)$ is the $xx$ component of 
vacuum Green's tensor describing the vectorial dipole-dipole interaction between atoms $i$ and $j$, 
including  the $1/r$, $1/r^2$ and $1/r^3$ terms~\cite{Ruostekoski1997,Novotny2006}.
As the trace of $[{\bf A}]$ is basis-independent,  
its imaginary part is the same expressed in the atomic basis as above or in 
the collective mode basis, hence $\sum_{m=1}^N\Gamma_m=N\Gamma_0$. 
In Fig.~\ref{Fig2}(b), we plot the normalized probability distribution of the spectral widths
extracted from  Fig.~\ref{Fig2}(a).  
We observe two peaks around $\Gamma=0$ and $\Gamma=2\Gamma_0$. 
They correspond to pairs of closely-spaced atoms for which the 
dipole moments are anti-parallel (subradiant) or parallel (superradiant), respectively. 
The maximum spectral width is  $\sim 5\Gamma_0$, showing 
that there is not a single mode dominating.

\begin{figure}
\includegraphics[width=\linewidth]{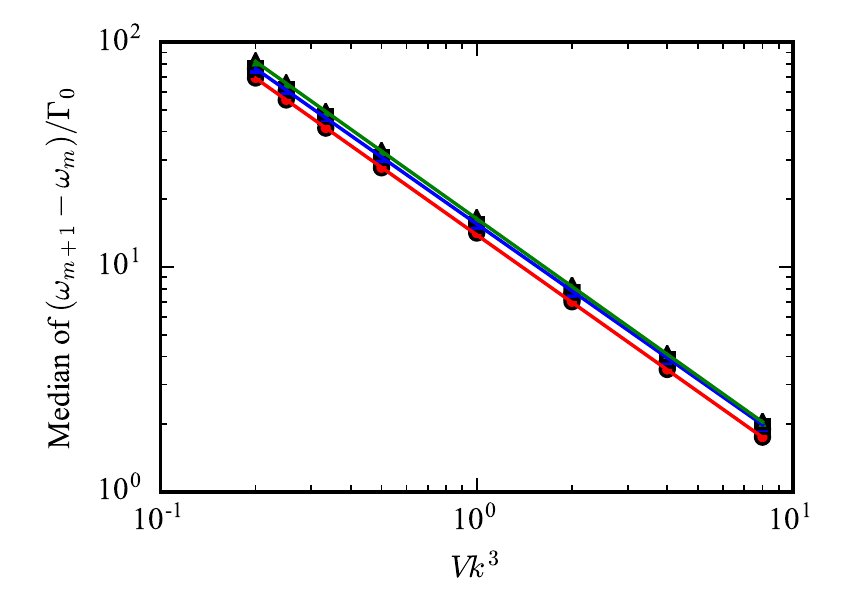}
\caption{Median of the spacing $\omega_{m+1}-\omega_m$ between two modes as a 
function of the volume of the box, for $N=20$ (red circles), $N=40$ (blue squares), 
and $N=80$ (green triangle). 
The numerical simulation follows the expected scaling 
$\delta\omega\sim\Gamma_0/(Vk^3)$ (lines).}
\label{Fig3}
\end{figure}

Having characterized the amplitude of the resonances and their width, we now investigate 
why the average scattered power does not depend on atom number for $N\gtrsim 20$. 
Let us define $\delta \omega$ as the typical frequency spacing between two eigenfrequencies. 
The probability that a monochromatic light with frequency $\omega$ 
excites a mode and scatters is thus $\sim\Gamma_0/\delta \omega$ so that 
the average scattering cross section of the 
cloud is approximately $\sigma\approx (\Gamma_0/\delta \omega)\,3\lambda^2/(2\pi) $. 
To estimate this average spacing $\delta \omega$, 
we first estimate the spread  $\Delta\omega$ of the eigenfrequencies. 
As the atomic ensemble is dense, the dipoles are mostly in the near-field of each other, 
dominated by the $1/r^3$ term. We thus assume that the frequency 
spread is given by the spectral shift $\Gamma_0/(kr)^3$ due to the interaction between two atoms 
separated by  a typical distance $r$ given  by $N r^3=V$~\cite{footnote_atom_pairs}. 
The spacing between modes is $\delta\omega\sim\Delta\omega/N$, 
where the atom number $N$ is also the total number of modes.
It follows that
\beq\label{Eq:delta_omega}
\delta\omega\sim\frac{\Gamma_0}{Vk^3}
\eeq
is independent of the number of atoms, and so is the average cross section. 
This scaling argument predicts a strong dependence of the frequency spacing with the volume. 
To check this prediction, we calculate the eigenmodes for a fixed number of atoms ($N= 20, 40, 80$)
while varying $Vk^3$ between $0.1$ and $10$.  
For each realization of a particular system we calculate all eigenfrequencies $\omega_m$. 
We then compute the spacings between adjacent modes $\omega_{m+1}-\omega_m$ 
and determine the median of this distribution as an indication of the typical spacing. 
We choose the median rather than the mean since it is insensitive to large spectral 
shifts originating from closely spaced pairs of atoms. 
The result is shown in Fig.~\ref{Fig3}: the mode spacing is indeed
independent of the atom number and is  proportional to $1/V$, 
thus confirming the  scaling of Eq.~(\ref{Eq:delta_omega}). 
It becomes larger than the average spectral width $\Gamma_0$ 
when $Vk^3\lesssim10$. 
Furthermore, as the widths of the modes are bounded by $N\Gamma_0$, 
for small enough volumes the modes do not overlap, their spacing being larger than their width. 
In summary, the strong dipole-dipole coupling between atoms produces collective modes 
spectrally well separated with a frequency spacing that varies as $1/(Vk^3)$. 
This near-resonance scattering regime differs considerably from the one of a 
sequence of scattering events by independent atoms, valid for weak interactions. 
The calculations above indicate that the transition between these two regimes occurs for 
$Vk^3\sim10$.
The factor controlling this transition is thus $Vk^3$, independent of $N$ provided $N\gtrsim 20$. 
Interestingly the atomic density at which the transition occurs 
is still orders of magnitude smaller than the densities encountered 
in condensed matter systems such as dielectric sphere or nano-particles.

\begin{figure}
\includegraphics[width=\linewidth]{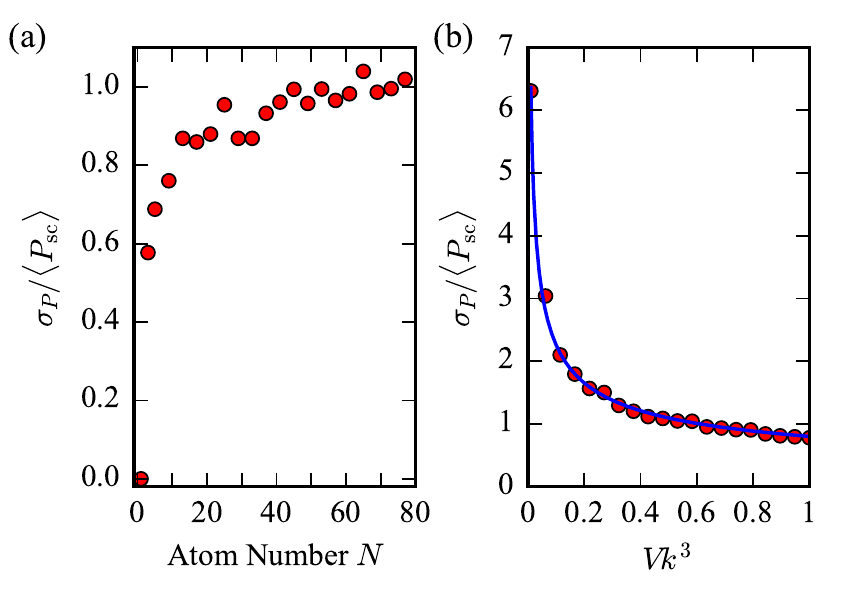}
\caption{(a) Standard deviation of the total scattered power normalized by 
the average scattered power, as a function of the atom number $N$ in a volume
$Vk^3=1$ for $\omega = \omega_0$. 
(b) Same quantity as a function of $Vk^3$ for $N=10$ and $\omega = \omega_0$. 
As $N$ increases, the standard deviation tends to an asympotic value scaling as   $\sim1/\sqrt{Vk^3}$: 
a fit gives $\sigma_P/\langle P_{\rm sc}\rangle= (0.80\pm0.03) (Vk^3)^{-(0.45\pm0.01)}$. 
}
\label{Fig4}
\end{figure}

In an actual experiment with a vapor, 
the atoms move so that the spectral positions of the modes fluctuate over time. 
For a monochromatic incident laser, we thus expect the scattered power to display giant fluctuations 
in time as the scattering cross section  fluctuates between $0$ and $3\lambda^2/(2\pi)$ (see Fig.~\ref{Fig1}b). 
To characterize the amplitude of the fluctuations, 
we compute the standard deviation of the scattered power 
over an ensemble of random realizations of the system: 
$\sigma_P=\sqrt{\langle P^2_\text{sc}\rangle-\langle P_\text{sc}\rangle^2}$.  
Figure~\ref{Fig4}(a) shows the standard deviation normalized 
by the average scattered power as a function of $N$: it 
becomes independent of the number of atoms as $N$ increases. 
We can understand this behavior using the following  argument. 
Assuming that the light is either scattered with a probability $\Gamma_0/\delta \omega$ or 
transmitted (i.e. not scattered) with a probability $1-\Gamma_0/\delta \omega$, we find that 
$\langle P_{\rm sc}\rangle = (\Gamma_0/\delta \omega)\sigma_0 I$ 
and $\langle P_{\rm sc}^2\rangle = (\Gamma_0/\delta \omega)\sigma_0^2 I^2$, with $\sigma_0$ the 
cross section of a mode and $I$ the intensity of the incoming light field. 
We thus obtain $\sigma_P/
\langle P_\text{sc}\rangle=\sqrt{(\delta\omega-\Gamma_0)/\Gamma_0}
\approx \sqrt{1/(Vk^3)}$ where we used Eq.~\eqref{Eq:delta_omega}. 
We plot  $\sigma_{\rm P}$ calculated numerically as a function of $V$ in Fig.~\ref{Fig4}(b). 
We indeed observe that the  fluctuations are very large for small volumes 
and follow the scaling derived above. 
In contrast, when the volume increases,  $\sigma_P$
tends to zero. 
The transition between the fluctuating to non-fluctuating regime 
occurs when the frequency spacing $\delta\omega$ 
equals the average spectral width $\Gamma_0$. 
According to Fig.~\ref{Fig3} and Eq.~(\ref{Eq:delta_omega}), 
this corresponds to $Vk^3\approx10$. In summary, for {\it interacting} atoms, 
the condition $Vk^3\lesssim10$ defines the large fluctuation regime with a small average scattering cross section. 
Remarkably for {\it non-interacting} atoms, the same condition defines the regime of 
coherent interferences leading to an enhanced scattering cross section $N^2 \, 3\lambda^2/(2\pi)$. 

So far, we have considered a nearly resonant illumination with a monochromatic source having 
a spectral width smaller than $\Gamma_0$. We now consider the case of a broad spectrum illumination 
with a width larger than the spectral width $\Delta \omega=N\delta \omega$. 
From the sum rule discussed previously, it follows 
that the spectrally integrated  cross section is constant and does not fluctuate 
from one realization to another or when the atoms move. The integrated cross section is thus 
expected to be $N\,3\lambda^2/(2\pi)$ and not to fluctuate around this value. 
This is an example of a self-averaging procedure~\cite{Sheng2006}: 
the fluctuations of the scattered light are removed by integrating over the many eigen-frequencies. 
An analogous system consists of light scattered 
by a rough surface forming a speckle pattern, which is detected using a collection solid angle 
either smaller or larger than the angular aperture of a speckle grain. 
In the former case, the signal will display large fluctuations when changing the realization. 
In the latter case, it will be constant although it is a measurement performed on a single realization. 
In both cases, the self-averaging procedure stems from the increase of the number of channels 
(different frequencies or different angles) used to transmit electromagnetic power 
while using a single realization of the random system.
 
In conclusion, we have analyzed the near-resonance scattering of light by
an ensemble of atoms in a volume $V$ smaller than $10/k^3$. 
When this condition is fulfilled, all the light induced dipoles are strongly interacting. 
This interaction produces spectrally well separated collective modes. 
The system needs to be described in terms of these collective modes and the 
picture of a sequence of scattering events by each atom is no longer valid. 
The scattering properties of this type of systems are: 
(i) the number of atoms only influences the total spectral width of the cloud; 
(ii) for a laser on resonance, the average scattered power and the fluctuations do not 
depend on the number of atoms for $N\gtrsim 20$, but only on the volume of the system, and 
(iii) the smooth transition between the usual coherent behavior 
of non interacting atoms and the large fluctuation regime takes place for $k^3V\approx 10$.
This transition thus occurs  in systems still very dilute with respect to condensed matter ones.

\acknowledgements{We thank R. Carminati and P. Pillet for discussions, 
and Igor Ferrier-Barbut for comments on the manuscript. 
We acknowledge support from the Triangle de la Physique (COLISCINA project), 
the French National Research Agency (ANR) as part of the ``Investissements d'Avenir'' 
program (Labex PALM, ANR-10-LABX-0039), and R\'egion Ile-de-France (LISCOLEM project). 
N.J.~S. is supported by Triangle de la Physique and Universit\'e Paris-Sud. 
J.-J.~G. acknowledges support from Institut Universitaire de France and the 
SAFRAN-IOGS chair on Ultimate Photonics.}


\begin{thebibliography}{50}
\bibitem{Footnote:2lvl} This corresponds to the case of a two-level atom featuring a $\pi$-transition.
Note that the dipole-dipole interaction is vectorial even for two-level atoms.
\bibitem{footnote_atom_pairs} 
In doing so, we neglect the contribution of the pairs of closely spaced atoms, which can lead
to large eigenfrequencies for certain realizations of the random spatial distribution.
\end{thebibliography}
\end{document}